\documentclass[preprint, aps, pre, preprintnumbers,amsmath,amssymb,showpacs]{revtex4}

\usepackage[final]{graphicx}  

\newcounter{saveeqn}

\begin{document}

\title{
Controlling the stability transfer between oppositely traveling
waves and standing waves by inversion-symmetry-breaking
perturbations}

\author{A.~Pinter\footnote[1]{Electronic address: kontakt@alexander-pinter.de},
M.~L\"ucke, and Ch.~Hoffmann}
\affiliation{Institut f\"ur Theoretische Physik, Universit\"at des Saarlandes,
Postfach 151150, \\ D-66041 Saarbr\"ucken, Germany}

\date{\today}

\begin{abstract}
The effect of an externally applied flow on symmetry degenerated
waves propagating into opposite directions and standing waves that
exchange stability with the traveling waves via mixed states is
analyzed. Wave structures that consist of spiral vortices in the
counter rotating Taylor-Couette system are investigated by full
numerical simulations and explained quantitatively by amplitude
equations containing quintic coupling terms. The latter are
appropriate to describe the influence of inversion symmetry breaking
perturbations on many oscillatory instabilities with $O(2)$
symmetry.
\end{abstract}

\pacs{47.20.Ky, 47.54.-r, 47.32.-y}
 \maketitle


Many nonlinear structure forming systems that are driven out of
equilibrium show a transition to traveling waves (TWs) as a result
of an oscillatory instability \cite{CH93}. In the presence of
inversion symmetry in one or two spatial directions also a standing
wave (SW) solution bifurcates in addition to the symmetry
degenerate, oppositely propagating TWs at the same threshold.
Moreover, depending on the parameters one can have a stability
exchange between TWs and SWs as the driving rate varies. The
stability transfer is mediated by mixed patterns that establish in
the solution space a connection between a pure TW and a pure SW.

Here we investigate how externally tunable symmetry breaking
perturbations change stability, bifurcation properties, and the
spatiotemporal behavior of the afore mentioned structures. To be
concrete we investigate vortex waves in the annular gap between
counter rotating concentric cylinders of the Taylor-Couette system
\cite{T94,CI94}. To that end we performed full numerical simulations
of the Navier-Stokes equations with methods described in
\cite{HLP04}. We elucidate that and how these results can be
explained quantitatively by amplitude equations which contain
only quintic order terms. 

The perturbation is realized in our system by an externally enforced
axial through-flow that can easily be controlled experimentally.
However, our results concerning the influence of inversion symmetry
breaking perturbations on TWs, SWs, and on the mixed states are more
general: the quintic order amplitude equations with small symmetry
breaking terms apply to all kinds of $O(2)$ symmetric oscillatory
instabilities in the presence of inversion symmetry breaking
perturbations.

{\it Structures} --
Without symmetry breaking through-flow the following oscillatory
vortex structures occur at small driving \cite{DI84,PLH06}: ({\em
i}) Forward bifurcating TWs consisting of left handed spiral
vortices (L-SPI) or of right handed spiral vortices (R-SPI) that are
mirror images of each other. L-SPI (R-SPI) travel in the annulus
between the two cylinders into (opposite to) the direction of the
rotation frequency vector of the inner one, i.e., in our notation
upwards (downwards) \cite{HLP04}. ({\em ii}) Forward bifurcating SWs
that consist of an equal-amplitude nonlinear combination of L-SPI
and R-SPI that are called ribbons (RIBs) in the Taylor-Couette
literature \cite{TESM89,LPA03}. ({\em iii}) So-called cross-spirals
(CR-SPI), i.e., combinations of L-SPI and R-SPI with different
amplitudes. They provide a stability transferring connection between
TW and SW solution branches \cite{CI94,PLH06}. The vortex structures
({\em i})-({\em iii}) are axially and azimuthally periodic with wave
numbers $k=2\pi /\lambda$ and $M$, respectively, with $\lambda=1.3$
and $M=2$ throughout this paper. They rotate with characteristic
constant angular velocities as a whole into the same direction as
the inner cylinder \cite{HLP04}. Thereby they are forced to
propagate axially with the exception of RIB vortices which rotate
only but do not propagate.

\begin{figure}
\includegraphics[clip=true,width=16.7cm, angle=0]{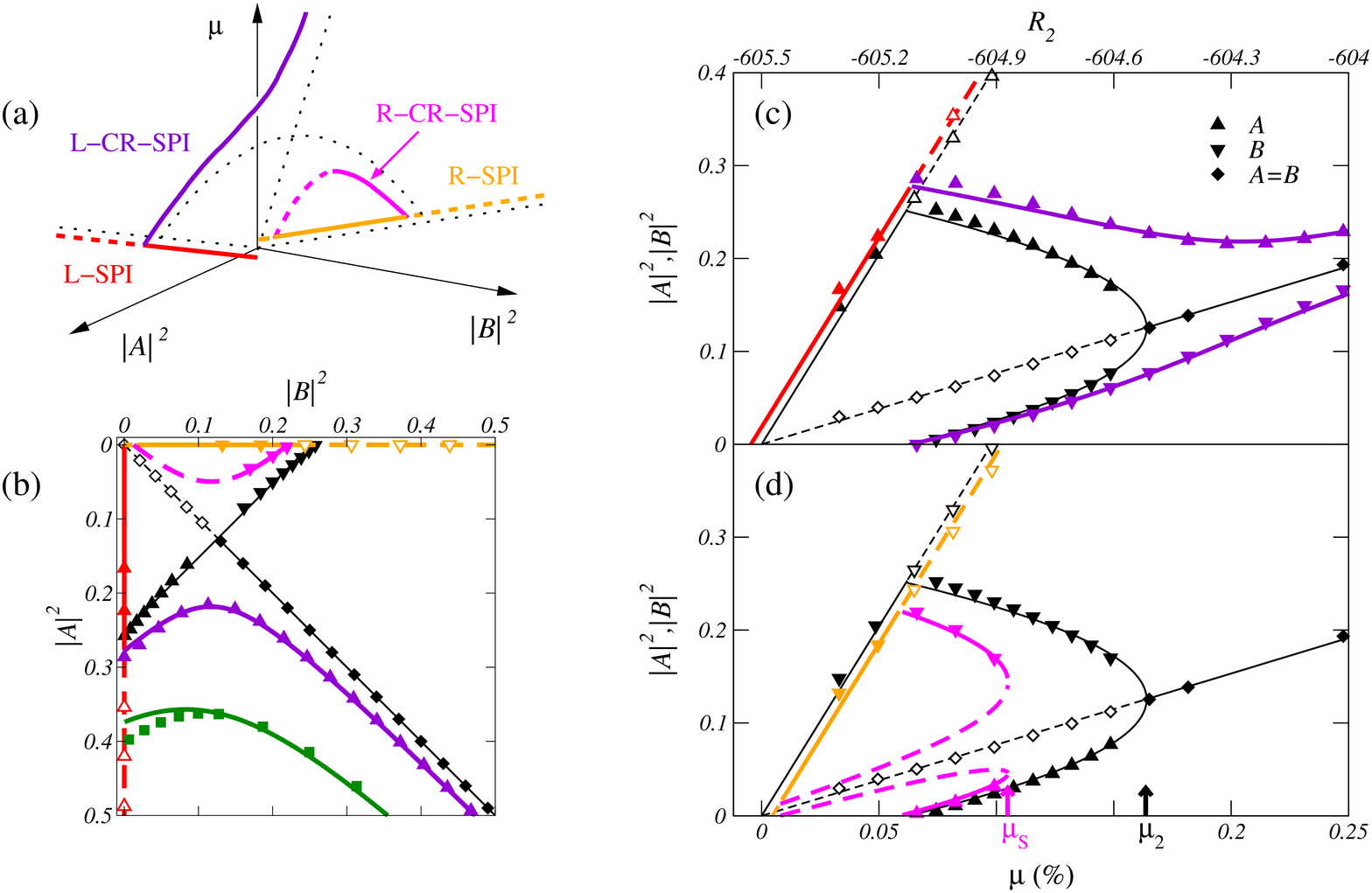}
\caption{\label{FIG-Bifurcation-Diagram}(Color) Bifurcation diagrams
of L-SPI $(A \neq 0, B=0, D>0)$, R-SPI $(A=0, B \neq 0, D<0)$,
L-CR-SPI $(|A|>|B|, D>0)$, R-CR-SPI $(|B|>|A|, D<0)$ and RIB (A=B,
D=0 -- only for $Re=0$) as functions of $\mu$ and $R_2$
for $R_1=240$. Here, $A$ and $B$ are the amplitudes of
the dominant modes $u_{2,1}(t)$ and $u_{2,-1}(t)$ (\ref{EQ:u2pm1}),
respectively, in the radial
velocity field $u$ at mid-gap. Solid (open) symbols denote stable
(unstable) solutions of the full Navier-Stokes equations. Full
(dashed) lines are the stable (unstable) solutions of the coupled
amplitude equations. Red, orange, violet, and magenta show
results for $Re=0.02$. The color coding of the different solutions
is given in the schematic 3D bifurcation diagram of (a). Therein the
black dots denote $Re=0$ solutions without stability information.
In (b) we show the projection of bifurcation diagrams onto the
$|A|^2-|B|^2$ plane. The color green refers to L-CR-SPI for $Re=0.1$
where R-CR-SPI do not exist anymore.  Black symbols and lines in (b)-(d) refer to $Re=0$.}
\end{figure}

\begin{figure}
\includegraphics[clip=true,width=8cm, angle=0]{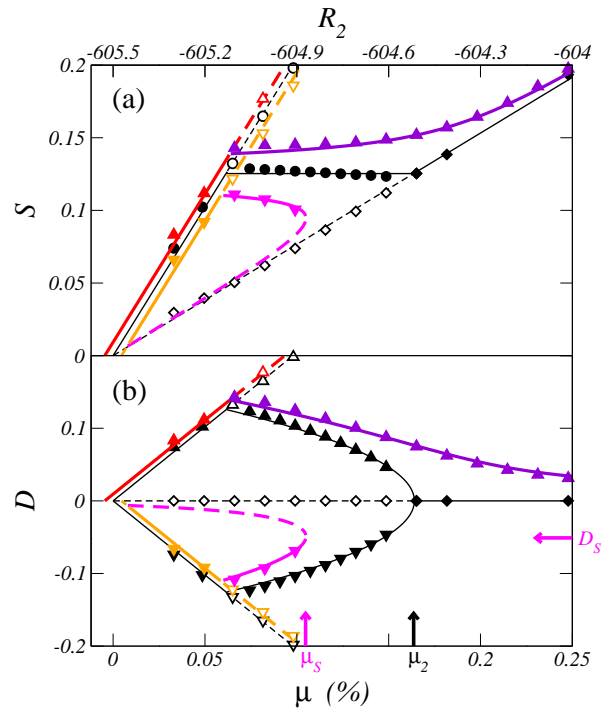}
\caption{\label{FIG-Bifurcation-Diagram-S-D}
(Color) Bifurcation behavior of
the vortex states of Fig.~\ref{FIG-Bifurcation-Diagram} are shown here in
plots of $S$ (a) and $D$ (b) versus $\mu$ and $R_2$, respectively. The color
coding is the same as in Fig.~\ref{FIG-Bifurcation-Diagram}.}
\end{figure}

{\it Order parameters} --
Figures
\ref{FIG-Bifurcation-Diagram}-\ref{FIG-Bifurcation-Diagram-S-D} show
bifurcation diagrams of SPI, CR-SPI, and RIB solutions in a system of radius ratio $\eta=1/2$ for fixed
$R_1=240$ versus the reduced distance
$\mu=\left(R_2-R_2^0\right)/\left|R_2^0\right|$
from the spiral onset $R_2^0$ in the absence of through-flow, $Re=0$. Here,
$R_1$ and $R_2$ are the Reynolds numbers defined by the rotational velocities
of the inner and outer cylinder, respectively, and $Re$ is the Reynolds number
of the imposed axial through-flow. In Figs. \ref{FIG-Bifurcation-Diagram} and
\ref{FIG-Bifurcation-Diagram-S-D}, the influence of a small through-flow
($Re=0.02$) is compared with the situation without through-flow. Order
parameters in Fig. \ref{FIG-Bifurcation-Diagram} are the squared amplitudes
$|A|^2,|B|^2$ of the dominant critical modes
\begin{equation} \label{EQ:u2pm1}
u_{2,1}(t)=|A|e^{-i\omega_A t}, \quad  \quad u_{2,-1}(t)=|B|e^{-i\omega_B t}\,
\end{equation}
in the double Fourier decomposition of the radial velocity $u$ at
mid-gap in azimuthal and axial direction. The indices 2 and $\pm 1$ identify azimuthal and axial modes, respectively. The linear stability analysis of the basic state consisting of a
superposition of circular Couette flow and of annular Poiseuille flow in axial
direction shows that the growth rates of these modes become
positive at the respective bifurcation thresholds $\mu_A$ and $\mu_B$ of $M=2$
L-SPI and R-SPI, respectively. Note also that for all relaxed vortex
structures investigated here the moduli and frequencies of $u_{2,\pm1}$ in
Eq.~(\ref{EQ:u2pm1}) are constant.

In Fig.~\ref{FIG-Bifurcation-Diagram-S-D} we show in addition the bifurcation
diagrams of the combinations
\begin{equation} \label{EQ:SD}
 S=\left(|A|^2+|B|^2\right)/2 ,\quad  \quad D=\left(|A|^2-|B|^2\right)/2\,
\end{equation}
since they are convenient to describe in particular CR-SPI.

{\it Bifurcation scenario for zero through-flow} --
In the symmetry degenerate case without through-flow L-SPI $(A \neq 0=B, D>0)$
and R-SPI $(A=0\neq B, D<0)$ have identical bifurcation properties. They are
stable close to onset whereas RIB $(A=B, D=0)$ are initially unstable. In the
driving range shown in figures \ref{FIG-Bifurcation-Diagram} and
\ref{FIG-Bifurcation-Diagram-S-D} the squared amplitudes of these two states
grow practically linearly with the reduced distance $\mu$ from the common
onset at $\mu=0$, albeit with different slopes. Then, there appear in a finite
supercritical driving interval stable CR-SPI solutions which transfer
stability from SPI to RIB. The solution which bifurcates with $B=0$ out of the
L-SPI is identified as a L-CR-SPI with $|A| > |B|$, i.e., $D>0$. The symmetry
degenerate R-CR-SPI $(|B|>|A|, D<0)$ bifurcates with $A=0$ out of the R-SPI.
In the former $|B|$ grows and $|A|$ decreases -- and vice versa in the latter
-- until the CR-SPI branches end with $A=B, D=0$ in the RIB solution. The
amplitude variations of the CR-SPI solutions, however, are such that the sum
$S$ remains practically constant, cf.
Fig.~\ref{FIG-Bifurcation-Diagram-S-D}(a). The RIB state loses stability
outside the plot range of figures \ref{FIG-Bifurcation-Diagram} and
\ref{FIG-Bifurcation-Diagram-S-D} to another type of amplitude-modulated
CR-SPI that are not discussed here.

{\it Through-flow induced changes} --
The axial through-flow significantly perturbs and changes structure, dynamics,
and bifurcation behavior of the SPI vortex solutions discussed so far \cite{HLP04,CK87,BP90,RL93,AMM06}: RIBs
cease to exist in the strict sense, L- and R-SPI are no longer mirror images
of each other, and also L-CR-SPI are no longer related to R-CR-SPI by this
symmetry operation. However, spirals retain their spatiotemporal structure in
the through-flow, i.e., they still do not depend on $\varphi,z,t$ separately
but only on the phase combination $\phi_A= M\varphi + kz-\omega_A t$ or
$\phi_B= M\varphi - kz-\omega_B t$, respectively. L-SPI (red color in Figs.
\ref{FIG-Bifurcation-Diagram}-\ref{FIG-Bifurcation-Diagram-S-D}) bifurcate
for small $Re>0$ at a threshold value $\mu_A<0$ prior to R-SPI (orange color)
which bifurcate at $0<\mu_B \approx -\mu_A$ out of the basic state.

L-SPI are again stable at onset, but then lose stability to L-CR-SPI (violet
color) which remain stable in the plotted
parameter regime. The L-CR-SPI solution approaches the $(Re=0)$ RIB state with
increasing $\mu$, but retains with $A\neq B$ a finite distance $D>0$. On the
other hand, R-SPI  are unstable for small and large $\mu$, but stable for
intermediate $\mu$. Stability is exchanged with a R-CR-SPI (magenta color ) which
has a stable as well as an unstable branch resulting from a
saddle-node bifurcation at $\mu_S$ in Figs.~\ref{FIG-Bifurcation-Diagram},
\ref{FIG-Bifurcation-Diagram-S-D}. For small $\mu$ the unstable R-CR-SPI lies
close to the $(Re=0)$ RIB solution and bifurcates with finite $D<0$ out of the
R-SPI slightly above $\mu_B$. Note also that the sum of the squared
amplitudes $S$ of CR-SPI is no longer constant as for the case of $Re=0$.

{\it Amplitude equations} --
The changes in spatiotemporal, bifurcation, and stability
behavior of SPI, CR-SPI, and RIB states by a small through-flow can be
explained and described close to onset quantitatively within an
amplitude-equation approach. To demonstrate that we focus here on
the bifurcation properties of the moduli $|A|, |B|$ of the critical modes.

In order to reproduce the bifurcation and stability behavior of the
aforementioned vortex states including the CR-SPI one needs coupled equations
for $A$ and $B$ of at least quintic order. Higher-order terms that are suggested in
\cite{CI94} are not necessary to ensure the existence of CR-SPI solutions.
Symmetry arguments \cite{PLH06,CI94,CK87} restrict the form of the equations for
the moduli to
\begin{subequations}\label{EQ-gekoppelte-AG}
\begin{eqnarray}
\tau_A\frac{d|A|}{dt}&=&|A|\left[\left(\mu-\mu_A\right)+b_A|A|^2+c_A|B|^2+e_A\left(|A|^2|B|^2-|B|^4\right)\right],\\
\tau_B\frac{d|B|}{dt}&=&|B|\left[\left(\mu-\mu_B\right)+b_B|B|^2+c_B|A|^2+e_B\left(|B|^2|A|^2-|A|^4\right)\right]
\end{eqnarray}
\end{subequations}
with real coefficients that depend in general on $Re$.
Here we have discarded the quintic terms $|A|^5$ and $|B|^5$ in view of the
linear variation of the squared SPI moduli with $\mu$, cf.
Fig.~\ref{FIG-Bifurcation-Diagram}. Furthermore, we made a special choice for
the coefficients of the last terms $|A||B|^4$ and $|B||A|^4$ in Eq.
(\ref{EQ-gekoppelte-AG}) that is motivated by the linear variation of the
squared RIB moduli with $\mu$ for $Re=0$ and that suffices to
describe the behavior for small $Re$ as well.

{\it Coefficients} --
As a result of the inversion symmetry under $z \leftrightarrow -z$ which
includes reverting the through-flow the coupled equations are invariant under
the operation $(A,B,Re) \leftrightarrow (B,A,-Re)$ so that the coefficients in
Eq.~(\ref{EQ-gekoppelte-AG}) obey relations like, e.g., $c_A(Re)=c_B(-Re)$. To
reproduce the bifurcation properties of the moduli for small $Re$ as in
Figs.~\ref{FIG-Bifurcation-Diagram}, \ref{FIG-Bifurcation-Diagram-S-D} it
suffices to incorporate the $Re$ dependence to linear order in the
coefficients
\begin{subequations}\label{EQ-coefficients}
\begin{eqnarray}
\mu_A(Re)=-\mu^{(1)} Re, && \quad \mu_B(Re)=\mu^{(1)} Re  \\
c_A(Re)=c+c^{(1)}Re, && \quad c_B(Re)=c-c^{(1)}Re
\end{eqnarray}
only and to ignore any $Re$ dependence of the others by setting
\begin{eqnarray}
b_A=b_B=b,\quad e_A=e_B=e.
\end{eqnarray}
\end{subequations}
The choice $b_A=b_B=b$ reflects the fact that the linear growth of the squared
SPI moduli with the distance from their respective thresholds at $\mu_A$ and
$\mu_B$ is unchanged by the through-flow. On the other hand,
the downwards shift of the L-SPI onset being for small through-flow of equal
magnitude as the upwards shift of the R-SPI onset is reflected by
$\mu_A(Re)=-\mu_B(Re)=-\mu^{(1)} Re$ with positive $\mu^{(1)}$. The coupling
constant $e_A=e_B=e$ ensures the existence of CR-SPI solutions \cite{PLH06}.
The flow induced changes of the coupling constants $c_A$ and $c_B$ reflect the
perturbation and destruction of the RIB states and ensures for positive $Re$
L-CR-SPI solutions with $D>0$ when $\mu$ is large. The values of the
coefficients \cite{coefficients} were obtained from linear stability analyses
and by fits to the full numerical nonlinear results.

{\it Fixed points} --
With the coefficients (\ref{EQ-coefficients}) it is straightforward to derive
from (\ref{EQ-gekoppelte-AG}) the following relations for the fixed points of
$S$ and $D$
\begin{subequations}\label{EQ-Fixpunkt-Re}
\begin{eqnarray}
\mu+\left(b+c\right)S- c^{(1)} Re D -2eD^2&=&0, \\
\mu^{(1)}Re +(b-c)D+c^{(1)}Re S +2eSD &=&0.
\end{eqnarray}
\end{subequations}
From these equations we have obtained the lines in the bifurcation plots of
$|A|^2=S+D$ and $|B|^2=S-D$ in Fig.~\ref{FIG-Bifurcation-Diagram} and of $S$
and $D$ in Fig.~\ref{FIG-Bifurcation-Diagram-S-D}. They compare very well with
the symbols from the full numerical simulations of the Navier-Stokes
equations, say, up to $Re\simeq 0.2$ or so. Beyond that higher order terms in
the $Re$-expansion of the coefficients should be included.

{\it Perturbation of the RIB $\to$ CR-SPI bifurcation} --
The black lines and symbols in Figs.~\ref{FIG-Bifurcation-Diagram},
\ref{FIG-Bifurcation-Diagram-S-D} show that without through-flow the RIB state
which is stable at large $\mu$ transfers stability to CR-SPI in a pitchfork
bifurcation located at $\mu_2$ in Fig.~\ref{FIG-Bifurcation-Diagram},
\ref{FIG-Bifurcation-Diagram-S-D}. The symmetry and topology of this
unperturbed bifurcation out of the $D=0$ RIB solution at
$\mu_2=-(b+c)S_{\ast}$, $S_{\ast}=(c-b)/(2e)$ is best seen in
Fig.~\ref{FIG-Bifurcation-Diagram-S-D}(b). This figure shows also most clearly
how the through-flow with $Re>0$ perturbs this bifurcation into a continuously
varying stable solution branch with $D>0$ (violet) and a pair of stable and unstable
solutions (magenta) that are connected by a saddle-node bifurcation at
$\mu_S$.

\begin{figure}
\includegraphics[clip=true,width=8cm, angle=0]{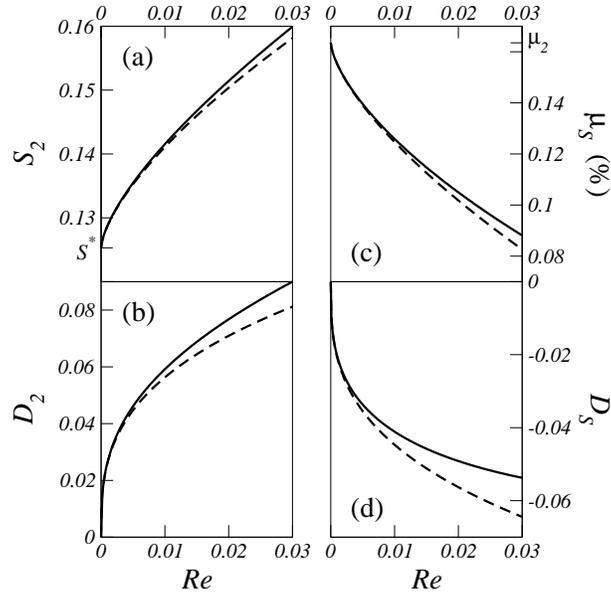}
\caption{\label{FIG-power-behavior}
Power law behavior of quantities
characterizing the break-up of the $Re=0$ pitchfork bifurcation from RIB to
CR-SPI versus through-flow strength $Re$ for $R_1=240$. $D_2$ and $S_2$ refer
to values at the location $\mu_2$ of the unperturbed bifurcation; $D_S$ and
$\mu_S$ are saddle node coordinates, cf.
Fig.~\ref{FIG-Bifurcation-Diagram-S-D}(b). Solid lines come from the full
fixed point equations (\ref{EQ-Fixpunkt-Re}); dashed lines refer to their
leading order approximations (\ref{EQ-S2-D2},\ref{EQ-muS-DS}).}
\end{figure}

The values of $S$ and $D$ at the unperturbed location $\mu=\mu_2$ of the
pitchfork bifurcation
\begin{equation}\label{EQ-S2-D2}
S(\mu_2)=S_2=S_{\ast}+\hat s_2Re^{2/3}+ h.o.t.,\qquad D(\mu_2)=D_2=\hat d_2Re^{1/3} + h.o.t.,
\end{equation}
grow $\propto Re^{2/3}$ and $\propto Re^{1/3}$, respectively, with constants
$\hat d^3_2=-(b+c)(\mu^{(1)} + c^{(1)} S_{\ast})/(4e^2)$ and $\hat s_2=2e \hat d_2^2/(b+c)$.
The coordinates of the saddle-node bifurcation, as indicated in
Fig.~\ref{FIG-Bifurcation-Diagram-S-D}(b) by the magenta arrows,
\begin{equation}\label{EQ-muS-DS}
\mu_S=\mu_2+\hat \mu_S Re^{2/3}+h.o.t., \qquad D_S=\hat d_S Re^{1/3}+h.o.t.,
\end{equation}
vary $\propto Re^{2/3}$ and $\propto Re^{1/3}$ with the perturbation strength
$Re$. Here the constants are
$\hat d^3_S=\left[(c^2-b^2)c^{(1)}+2e(b+c)\mu^{(1)}\right]/\left(16e^3\right)$ and
$\hat \mu_S=\left[(b+c)\mu^{(1)}-\mu_2c^{(1)}+4e^2 \hat d_S^3\right]/(2e\hat d_S)$. The
comparison in Fig.~\ref{FIG-power-behavior} of these lowest order
approximations (\ref{EQ-S2-D2}, \ref{EQ-muS-DS}) that reflect an additive pitchfork unfolding 
with the full fixed point result from Eq.~(\ref{EQ-Fixpunkt-Re}) shows good agreement for $Re \lesssim
0.02$.

\begin{figure}
\includegraphics[clip=true,width=8cm, angle=0]{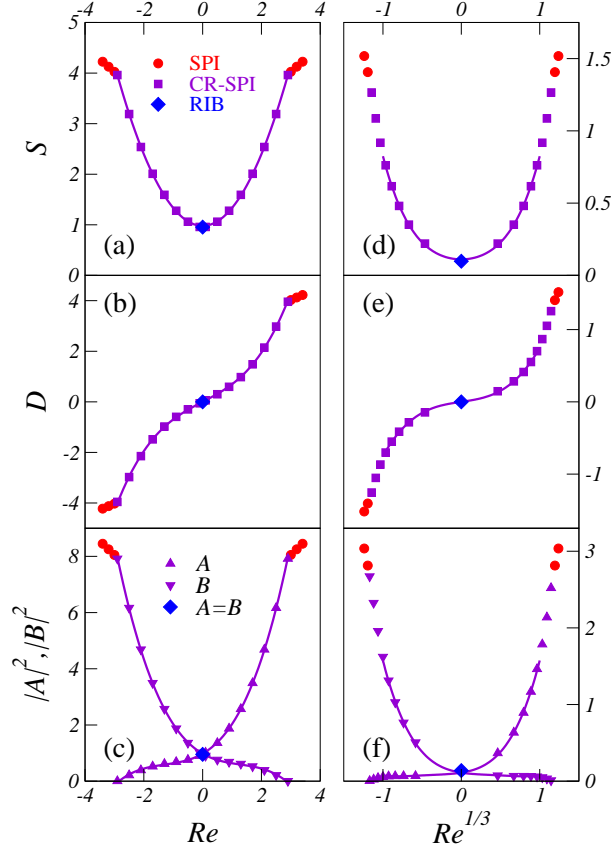}
\caption{\label{FIG-throughflow} (Color online) Power-law variation of $S,D,|A|^2=S+D,|B|^2=S-D$ with through-flow $Re$ during the transformation from RIB (diamond) at $Re=0$ to CR-SPI (squares and triangles). The latter undergo a transition to SPI (circles) at large $Re$. Symbols denote results from the full Navier-Stokes equations for $R_1=240$, lines show fits explained in the text. Right column: $\mu = \mu_2$, i.e., the unperturbed pitch fork location; left column $\mu=0.011$, i.e., far away from it. Note the different abscissa scaling of the two columns.}
\end{figure}

{\it Larger through-flow} --
Here we investigate the transformation of the unperturbed RIB state into
a CR-SPI as a function of the through-flow strength extending to larger Reynolds numbers up to $Re=4$ which are about 200 times larger than those studied so far in this paper. For such large $Re$ the amplitude equations with the simple $Re$ dependence of the coefficients (\ref{EQ-coefficients}) are
no longer adequate to describe the behavior of the order parameters $S$ and $D$.
However, they continue to follow the same simple power laws with $Re$ as the ones for small through-flow: When $\mu$ is well away from the location $\mu_2$ of the unperturbed pitchfork bifurcation as in Fig.~\ref{FIG-throughflow}(a-c) one has $S=S^{(0)}+S^{(2)}Re^2\,, D=D^{(1)}Re+D^{(3)}Re^3$. On the other hand, at $\mu_2$, as in Fig.~\ref{FIG-throughflow}(d-f) one has $S=S^{(0)}+S^{(2/3)}(Re^{1/3})^2+ S^{(4/3)}(Re^{1/3})^4\,, D=D^{(1/3)}Re^{1/3}+D^{(1)}(Re^{1/3})^3$. The lines in
Fig.~\ref{FIG-throughflow} show fits of the data from the full numerical simulations (symbols) to these power law expansions for $S$ and $D$. Fig.~\ref{FIG-throughflow}(c,f) shows plots of the resulting squared moduli $|A|^2=S+D$ and $|B|^2=S-D$.

Note that inversion symmetry $z \leftrightarrow -z$ implies $S(-Re)=S(Re)$ and $D(-Re)=-D(Re)$. For $Re>0$ $(Re<0)$ the RIB state is transformed into a L-CR-SPI (R-CR-SPI) with $A$ ($B$) being the major mode and $B$ ($A$) the minor one. With increasing through-flow the major (minor) mode of the CR-SPI increases (decreases) until the minor mode vanishes in the transition to a pure SPI.

{\it Conclusion} --
A small applied flow is an inversion symmetry breaking perturbation
for fluid wave structures that is easy to control. It significantly
changes the bifurcation and stability transfer scenario involving
symmetry degenerate TWs, mirror symmetric SWs, and mixed wave
patterns. Numerical simulations of such waves in the
counter-rotating Taylor-Couette system are explained and reproduced
quantitatively by amplitude equations containing quintic order
coupling and symmetry breaking perturbation terms. This approach can
be applied to a wide range of $O(2)$ symmetric oscillatory
instabilities in the presence of inversion symmetry breaking
perturbations.

This work was supported by the Deutsche Forschungsgemeinschaft.

\newpage

\end{document}